\documentclass[11pt,a4paper]{article}
\usepackage{jheppub}
\usepackage{amsmath}
\usepackage[most]{tcolorbox}
\usepackage{dsfont}
\usepackage{ulem}
\usepackage{natbib}
\usepackage{xcolor}
\usepackage[hang,flushmargin]{footmisc}
\usepackage{tikz-cd}
\usepackage{enumitem}
\usepackage{amsfonts,amssymb, amscd,amsmath,latexsym,amsbsy,bm}
\usepackage{stmaryrd}
\usepackage{todonotes}

\DeclareMathOperator*{\SumInt}{%
	\mathchoice%
	{\ooalign{$\displaystyle\sum$\cr\hidewidth$\displaystyle\int$\hidewidth\cr}}
	{\ooalign{\raisebox{.14\height}{\scalebox{.7}{$\textstyle\sum$}}\cr\hidewidth$\textstyle\int$\hidewidth\cr}}
	{\ooalign{\raisebox{.2\height}{\scalebox{.6}{$\scriptstyle\sum$}}\cr$\scriptstyle\int$\cr}}
	{\ooalign{\raisebox{.2\height}{\scalebox{.6}{$\scriptstyle\sum$}}\cr$\scriptstyle\int$\cr}}
}

\makeatletter\renewcommand{\@biblabel}[1]{#1.}\makeatother

\newtcolorbox{empheqboxed}{colback=gray!20, 
	colframe=white,
	width=\textwidth,
	sharpish corners,
	top=0mm, 
	bottom=0pt
}

\title{Integrability from supersymmetric duality: a short review}

\author{Ilmar Gahramanov$^{a,b,c}$}
\affiliation{$^a$ Department of Physics, Bogazici University,\\ 34342 Bebek, Istanbul, Turkey\\[-0.5cm]
	
	$^b$ Department of Mathematics, Khazar University, \\ Mehseti St. 41, AZ1096, Baku, Azerbaijan\\[-0.5cm]
	
	$^{c}$ Institute of Radiation Problems, Azerbaijan National Academy of Sciences, \\ B.Vahabzade St. 9, AZ1143, Baku, Azerbaijan\\[-0.5cm]
}

\emailAdd{ilmar.gahramanov@boun.edu.tr}

\abstract{ Integrable models of statistical mechanics play a prominent role in modern mathematical physics, especially in conformal field theory, knot theory, combinatorics, topology, etc. In this brief review, we discuss a program of constructing integrable lattice spin models with the nearest neighbor interaction using methods inspired by the supersymmetric gauge theory computations, called gauge/YBE correspondence. After a brief introduction to the topic, we review some recent examples of this correspondence and the role of special functions and symmetries. Finally, we discuss future directions of research.}


\begin{document}
	\maketitle
	\flushbottom

	\section{Introduction}
	
	There is a wealth of research involving lattice models of statistical mechanics. In physics, they mainly play a role as models for studying phase transitions \cite{baxter2007exactly}. The reason for an active study of such models comes also from an attempt to understand connections between this subject and other areas of mathematical physics such as special functions, enumerative combinatorics, graph theory, number theory, knot theory, quantum groups, etc.

	Recently there was observed a relationship between exact results in supersymmetric quiver gauge theories and exactly solvable two-dimensional lattice models in statistical mechanics, called the gauge/YBE correspondence. Due to this correspondence, the integrability in statistical models is a direct consequence of supersymmetric IR duality. Namely, it relates the Yang-Baxter equation with the equality of partition functions for supersymmetric dual theories. This relationship has led to the construction of new integrable models of statistical mechanics and we believe that much more is to be found.
	
	The success of supersymmetric gauge theory computations for integrable models stimulated the study of the relevant underlying mathematical structure. The objective of this paper is twofold, on one hand, it is intended as a short introduction to the gauge/YBE correspondence, on the other hand along the way we will discuss some mathematical issues and open problems, such as the role of special functions and symmetries in the study of integrable models. These notes are far from a comprehensive review, we do not provide all aspects of the progress in this area.

	The structure of this article is organized as follows. In sections 2 and 3 we briefly discuss some symmetry aspects of integrable models. We begin in the next section by considering the Bazhanov-Sergeev model. Finally, Section 4 discusses some possible directions to be faced by studies of integrable lattice models in the future. We will briefly discuss the recently observed connection relating supersymmetric gauge theories to integrable lattice models.

	\subsection{Integrable lattice models}
	
	Let us briefly review the integrable lattice models. The point of departure is the Yang-Baxter equation \cite{Takhtajan:1979iv,baxter1986yang,Jimbo:1989qm,Jimbo:1989mc}. The study of solutions to this equation has led to major breakthroughs in many areas of mathematical and theoretical physics\footnote{Besides theoretical studies of the Yang-Baxter equation it has also some experimental implementations, see, e.g. \cite{Batchelor:2015osa,Vind:2016yuh}.}. In integrable models of statistical mechanics, the Yang-Baxter equation implies the commutativity of transfer matrices, i.e. it is a sufficient condition for integrability. For our purpose we will consider the simplest type of lattice model, the so-called the Ising-type models\footnote{It has several names in the literature, as ``edge-interacting model'' or just ``spin-edge model''.} with local Boltzmann weights (for more details, see e.g. \cite{baxter2007exactly,baxter2010some,Bax02rip,Bazhanov:2016ajm}). In a few words, the Ising-type lattice is defined on a square lattice (or on any planar graph) with pairwise edge interaction\footnote{There are also vertex models where spins live on edges and interact at a vertex and IRF models where spins live on sites and interact round a face.} depending on spin variables situated at sites of a lattice and having discrete or/and continuous values\footnote{In the case of the well-known Ising model spins have only two values $+1$ and $-1$.}. In order to describe properties of the system we use the canonical ensemble and introduce the partition function\footnote{In this paper we will use the same notations as in \cite{Gahramanov:2017ysd}.} 
	\begin{equation}
		Z= \SumInt_{\{\sigma\}} e^{-\beta E(\sigma)} \;,
	\end{equation}
	where $\beta$ is the inverse temperature and $E$ stands for the total energy of the system $E=\sum_{(ij)} \epsilon(\sigma_i,\sigma_j)$ where the summation is over all edges $(ij)$ of the lattice and $\epsilon(\sigma_i,\sigma_j)$ is the energy. It is usual to use the Boltzmann weight notation instead of the energy in lattice spin models. One introduces the Boltzmann weight associated with the edge $(i,j)$
	\begin{equation}
		W(\sigma_i, \sigma_j) \ = \ e^{-\beta \epsilon(\sigma_i, \sigma_j)} \;.
	\end{equation}

	\begin{figure}[tbp]
		\centering 
		\includegraphics[width=.65\textwidth,trim=0 180 0 100,clip]{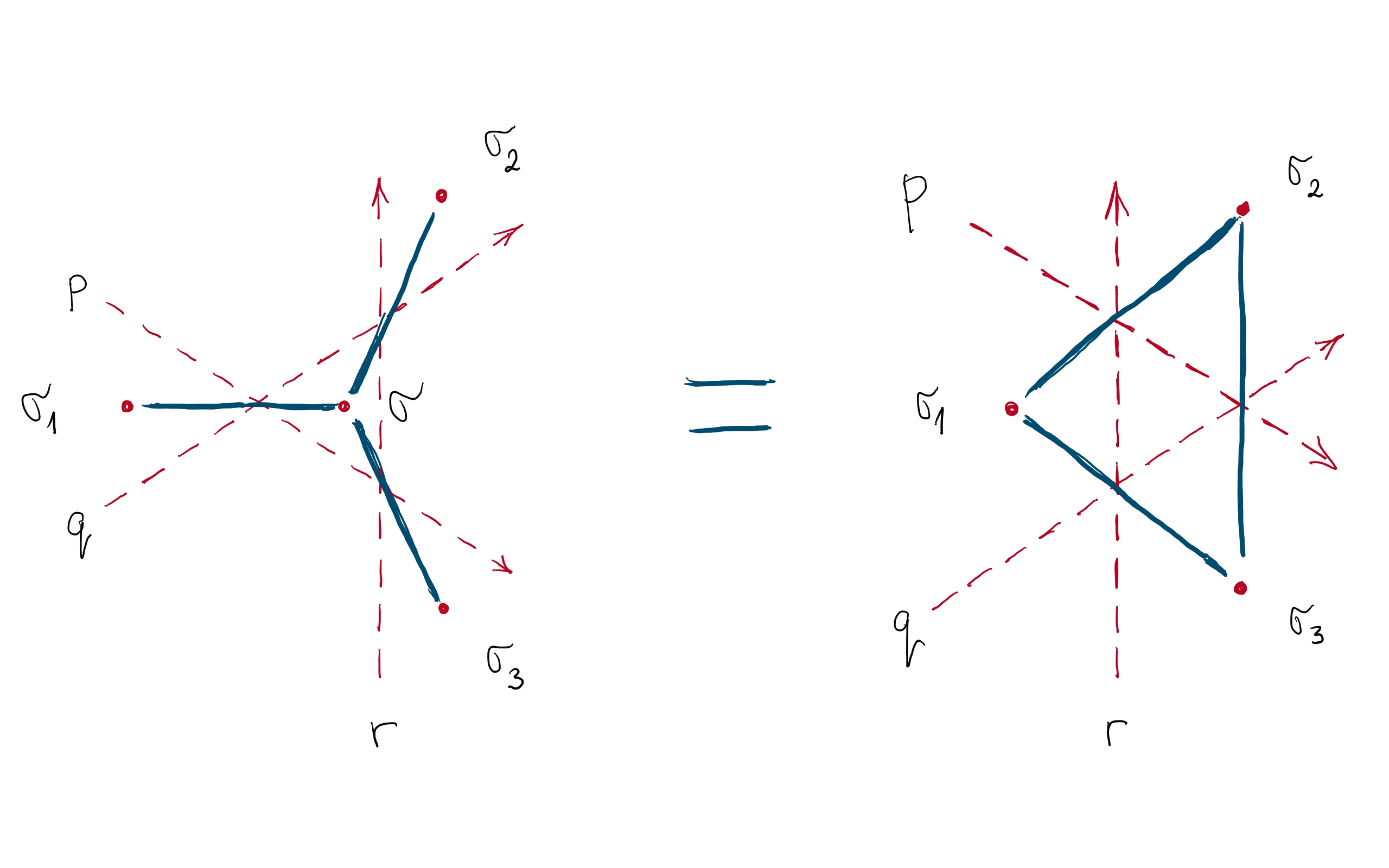}
		\caption{\label{a}  Graphical representation of the star-triangle relation: there is a ``star'' on the left side and ``triangle'' on the right side.}
	\end{figure}
	
	In all integrable lattice models the Boltzmann weights depend also on rapidity variables $p, q$. Morevover, for simplicity we suppose that our models have rapidity difference property, i.e. Boltzmann weights must depend only on the difference $p-q :=\alpha$. Then the partition function in terms of Boltzmann weights  has the following form
	\begin{equation}
		Z= \SumInt_{\{\sigma\}} \prod_{(i,j)} W_\alpha(\sigma_i, \sigma_j) \;,
	\end{equation}
	where the sum/integral is over all values of spins and the product is over all nearest-neighbour pairs $(i,j)$ of spins $\sigma_i, \; \sigma_j$. In some cases the partition function in the thermodynamic limit can be solved exactly, i.e.  computed without any approximation. We call such systems integrable or exactly solvable models \cite{baxter2007exactly}. There exists a sufficient condition for integrability\footnote{Under certain conditions the star-triangle relation represents also a necessary condition for integrability, see, e.g. \cite{Lochak:1984pe}.} of a system which for two-dimensional statistical mechanics on lattices is called the star-triangle relation \cite{Baxter:1997tn}
	\begin{align} \nonumber
		& \SumInt d\sigma \; S(\sigma) W_{\eta -\alpha}(\sigma_1, \sigma) W_{\eta -\beta}(\sigma_2, \sigma)  W_{\eta -\gamma}(\sigma_3, \sigma) \\
		& \qquad \qquad = \ R(\alpha, \beta, \gamma) W_{\alpha}(\sigma_2, \sigma_3) W_{\beta}(\sigma_1, \sigma_3) W_{\gamma}(\sigma_1, \sigma_2) \;,
	\end{align}
	where the Boltzmann weight $W(\sigma_i, \sigma_j)$ denotes an interaction between neighbouring spin pairs $\sigma_i$ and $\sigma_j$, $S(\sigma)$ stands for a self-interaction,  $\alpha, \beta, \gamma$ are rapidity parameters with the condition $\alpha+\beta+\gamma=\eta$ and $R(\alpha, \beta, \gamma)$ is some function symmetric in rapidity parameters. The star-triangle relation is a special form of the Yang-Baxter equation.
	
	In general, it is a very complicated problem to find statistical mechanical models satisfying the star-triangle relation.

	\subsection{Exact results in supersymmetric gauge theories}

	In recent years, there has been significant progress in the study of supersymmetric gauge theories in diverse dimensions thanks to the use of the localization technique \cite{Pestun:2007rz}. This technique enables us to compute exact quantities such as supersymmetric indices, partition functions, Wilson loops, ’t Hooft loops, surface operators, etc. on compact manifolds. The interested reader is referred to the review papers \cite{Hosomichi:2015jta,Cremonesi:2014dva,Pestun:2016jze,Pestun:2016zxk} for details on the subject.
	
	Our main interest will be the supersymmetric partition function on a manifold\footnote{The result, of course, depends on the choice of a manifold.} $M$
	\begin{equation} \label{eucpart}
		Z \ = \ \int [D\phi] \; e^{-S[\phi]} \; ,
	\end{equation}
	where $\phi$ denotes all the fields in a theory.  The supersymetric localization technique \cite{Pestun:2007rz} allows us to compute the partition function (\ref{eucpart}) exactly. Assume we consider a supersymmetric gauge theory on a manifold $M$ with a supercharge\footnote{In this paper we will consider supersymmetric gauge theories only with four supercharges.} $Q$ which is symmetry of the action, i.e. $QS \ = \ 0$. The key point is that one can use the fermionic operator $Q$ and modife the path-integral by a $Q$-exact term $tQV$ in the following way
	\begin{equation} \label{partfunct}
		Z(t) \ = \ \int [D \phi] e^{-S[\phi]-tQ V[\phi]} \; ,
	\end{equation}
	with $\delta_Q V=0$, where $V$ is some functional of fields and $\delta_Q$ is a supersymmetric transformation\footnote{The square of a supercharge (or any fermionic operator) is either zero or a bosonic symmetry $\delta_Q$ of the action.}.  If the measure is $Q$-invariant the partition function (\ref{partfunct}) is independent of parameter $t$ and one can use satisfying a particular saddle-point procedure. Therefore expanding fields near saddle points as $\phi=\phi_0+t^{-1/2} \phi'$ and taking the large $t$ expansion the path integral only gets contributions near $\delta_Q V[\phi_0]=0$ (BPS field configurations).  One ends up with a finite-dimensional integral for a matrix model which is exact and can be written in a closed form. 
	\begin{equation}
		Z \ = \ \int D \phi_0 \; e^{-S[\phi_0]} \; Z_{{1-loop}}[\phi_0] \;. 
	\end{equation}
	
	In the paper we mainly use the four-dimensional $\mathcal N=1$ superconformal index which is generalization of the Witten index \cite{Witten:1982df} by including global symmetries of a theory commuting with a particular supercharge \cite{Kinney:2005ej, Romelsberger:2005eg, Romelsberger:2007ec}
	\begin{equation}
		\text{Tr} (-1)^F e^{-\beta \{ Q, Q^{\dagger}\}} \prod t_i^{F_i} \;,
	\end{equation}
	where the trace is taken over the Hilbert space on $S^3$, $F_i$ are generators for global symmetries that commute with $Q$ and $Q^{\dagger}$, and $t_i$ are additional regulators corresponding to the global symmetries. One can interpret the superconformal index as a twisted partition function on a manifold $S_b^3 \times S^1$, where $S_b^3$ is a squashed three-sphere with the squashing parameter $b$ and use the supersymmetric localization technique for computing the index. In this case it can be presented as the following matrix integral
	\begin{equation}
		Z = \frac{1}{|W|} \int \prod_{i=1}^{rank \; G} dz_i \; Z_{vec} \; Z_{chiral} \;,
	\end{equation}
	where $z_i$ parametrize the maximal torus of gauge group $G$, $|W|$ is the order of the Weyl group, and $Z_{vec}, Z_{chiral}$ denote the contribution of vector and chiral multiplets, respectively.

	\subsection{Beta integrals and gauge/YBE correspondence}

	The gauge/YBE correspondence gives a precise method to construct solutions to the star-triangle equation by applying the exact supersymmetry computations\footnote{We include only those references that would help to understand the main points in the subject, a more complete list of references can be found in the review paper \cite{Gahramanov:2017ysd}.}. Equalities of supersymmetric partition functions of supersymmetric dual theories (such as Seiberg duality, Seiberg-like duality, mirror symmetry, etc.) lead to complicated integral identities for the various type of hypergeometric integrals \cite{Dolan:2008qi,Spiridonov:2009za,Dolan:2011rp,Spiridonov:2011hf,Gahramanov:gka,Gahramanov:2015tta,Gahramanov:2016wxi,Gahramanov:2014ona,Rosengren:2016mnw,Bozkurt:2018xno,Eren:2019ibl,Spiridonov:2019kto,Bozkurt:2020gyy,Catak:2021coz,Gahramanov:2021pgu}. Actually, the construction of integrable lattice spin models of statistical mechanics via the so-called gauge/YBE correspondence is based on such integral identities obtained mainly by using supersymmetric gauge theory results (see, e.g. reviews on the subject \cite{Gahramanov:2017ysd,Yamazaki:2018xbx}). We should note that this correspondence is quite a technical field, so one needs to be familiar with hypergeometric integrals and their properties. As we will see, by using this correspondence the search for
	a solution to the Yang-Baxter equation is easier to handle.

	To illustrate the concept of the gauge/YBE correspondence in a simple setting, we pick one particular example:  First let us recall the definition of the elliptic gamma function introduced by Ruijsenaars in \cite{Ruijsenaars:1997:FOA} and studied extensively in \cite{felder2000elliptic, Felder:2000mq}. It is the following meromorphic function of three complex variables $p,q$ and $z$ \cite{Ruijsenaars:1997:FOA}
	\begin{equation}
		\Gamma(z,p,q) = \prod_{i,j=0}^{\infty} \frac{1-z^{-1} p^{i+1} q^{j+1}}{1-z p^i q^j}, \; \; ~~~~ \text{with $|p|,|q|<1$.}
	\end{equation} 
	We will need also the q-Pochhammer symbol, defined as  the following infinite product
	\begin{equation}
		(z,q)_{\infty} = \prod_{i=0}^{\infty} (1- z q^i) \;.
	\end{equation}
	Now we can write the elliptic beta integral\footnote{There are many types of beta integrals, the
		interested reader is referred to \cite{gasper1989q,askey1986more,neretin2015matrix} and to the review on the role of such integrals in supersymmetric gauge theories \cite{Gahramanov:2015tta}.} introduced by Spiridonov \cite{spiridonov2001elliptic}
	\begin{equation} \label{betaint}
		\frac{(p;p)_\infty (q;q)_\infty}{2} \int_{\mathbb{T}} \frac{\prod_{i=1}^6 \Gamma(t_i z ;p,q)\Gamma(t_i z^{-1} ;p,q)}{\Gamma(z^{2};p,q) \Gamma(z^{-2};p,q)} \frac{dz}{2 \pi \textup{i} z} = \prod_{1 \leq i < j \leq 6} \Gamma(t_i t_j;p,q),
	\end{equation}
	where $t_1, \dots ,t_6,p,q \in {\mathbb{C}}$ with $|t_1|, \dots , |t_6|,|p|,|q| <1$, the unit circle $\mathbb{T}$ is taken in the positive orientation and we imposed the balancing condition $\prod_{i=1}^6 t_i=pq$. 
 
	The integral is an example of hyergeometric integral of elliptic type\footnote{The first example of the elliptic hypergeometric series was discovered about 25 years ago by Frenkel and Turaev \cite{frenkel1997elliptic} in the context of elliptic $6j$-symbol \cite{biedenharn1981angular}. This family of functions is the top level of hypergeometric functions \cite{Spiridonov:2010yc,Spiridonov:2016bxc}. Recently they have attracted attention of physicists since they proved to be a useful tool in theoretical and mathematical physics. An introduction to elliptic hypergeometric functions can be found, e.g. in \cite{Spiridonovessay,Spiridonov:2013zma,Gahramanov:2015tta,Rosengren:2016qtr,Rosengren:2017ujl,Spiridonov:2019kto,Spiridonov:2019dov}}.
	
	\begin{figure}[tbp]
		\centering 
		\includegraphics[width=.85\textwidth,trim=0 180 0 100,clip]{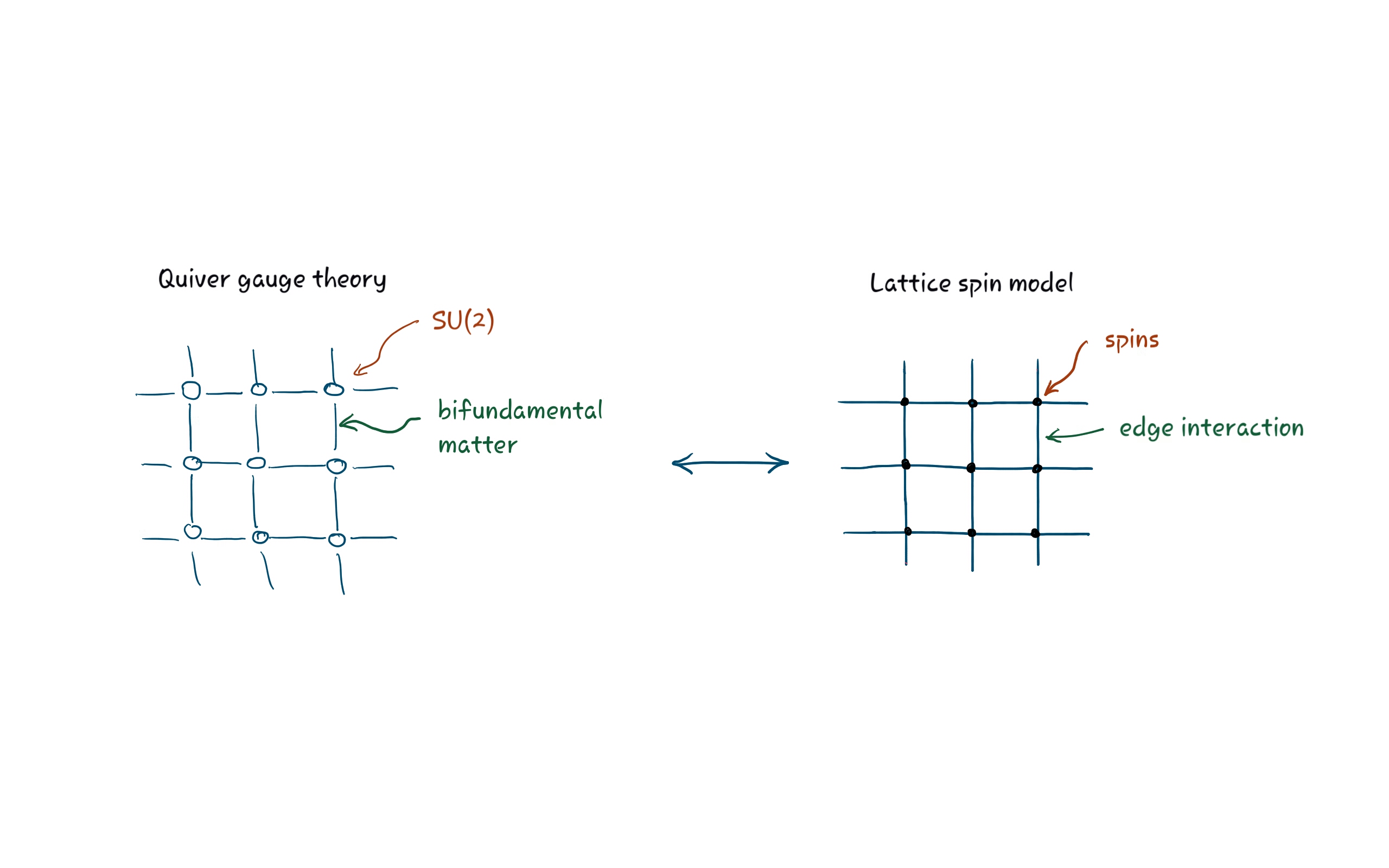}
		\caption{\label{ccc} Graphical representation of the gauge/YBE correspondence.}
	\end{figure}

	It turns out that the integral identity (\ref{betaint}) can be understood as a result of the special type of Seiberg duality\footnote{In the 1990s Seiberg \cite{Seiberg:1994pq} has shown a non-trivial quantum equivalence between different supersymmetric theories, called supersymmetric duality which means that two or more different theories may describe the same physics in the far-infrared limit.}  for the following four-dimensional $\mathcal N=1$ supersymmetric gauge theories \footnote{The computation can be done by techniques of supersymmetry localization (see, e.g. \cite{Pestun:2016zxk}), as well as by counting all the possible operators with certain weights and using the plethystic exponential (see, e.g. \cite{Romelsberger:2005eg,Romelsberger:2007ec,Gahramanov:2016sen}). } \cite{Dolan:2008qi}
	\begin{itemize}
		\item The left-hand side of the identity represents\footnote{The precise expression of the supersymmetric index depends on the details of the assignments of $U(1)_R$ charges for chiral multiplet.} the supersymmetric index of a theory with $SU(2)$ gauge group, chiral multiplets in the six-dimensional fundamental representation of $SU(6)$ flavor symmetry.
		
		\item The right-hand side stands for the superconformal index of the theory without gauge degrees of freedom, meson fields in the fifteen-dimensional antisymmetric tensor representation of $SU(6)$ flavor group.
	\end{itemize}
	Since these theories are dual, i.e. show the same physics at the infrared fixed point their superconformal indices \cite{Romelsberger:2005eg,Romelsberger:2007ec} should be equal and the expression (\ref{betaint}) represents this equality\footnote{The entry of elliptic hypergeometric integral identities into high energy physics occurred in 2008 when Dolan and Osborn observed [25] that the superconformal index can be expressed in terms of elliptic hypergeometric integrals.}. 
	
	Bazhanov and Sergeev showed that the beta integral can be written in the form of the star-triangle relation, i.e. the Boltzmann weights and R-factor are expressed in terms of elliptic gamma functions \cite{Bazhanov:2010kz}
	\begin{align}
		W_{\alpha}(\sigma_i,\sigma_j) & = \Gamma(e^{\alpha-\eta\pm i\sigma_i\pm i\sigma_j};p,q)\Gamma(e^{\alpha-\eta + i\sigma_i- i\sigma_j};p,q)\Gamma(e^{\alpha-\eta- i\sigma_i+ i\sigma_j};p,q)\Gamma(e^{\alpha-\eta- i\sigma_i- i\sigma_j};p,q) \;,
	\end{align}
	\begin{equation}
		S(\sigma)  =  \frac{(p;p)_ \infty(q;q)_ \infty  }{4 \pi } \frac{1}{\Gamma(e^{ - 2 i \sigma},p,q) \Gamma(e^{2 i \sigma},p,q)} \;,
	\end{equation}
	\begin{equation}
		R(\alpha, \beta, \gamma) \ = \  \Gamma(e^{-\alpha}; p, q) \Gamma(e^{-\beta}; p, q) \Gamma(e^{-\gamma}; p, q) \;,
	\end{equation}
	where $\alpha, \beta, \gamma$ stand for rapidity variables and $\eta=p q$ is a crossing parameter. 
	
	Since the same integral identity appears in two different areas, it is natural to look for a correspondence between this subjects  \cite{Spiridonov:2010em}. As one can see, it is not a hard task to obtain the star-triangle relation from the identity of superconformal indices. By adding a certain superpotential one breaks $SU(6)$ flavor symmetry to $SU(2) \times SU(2) \times SU(2)$ and use the general $R$-charges, namely one redefines fugacities in the identity (\ref{betaint}) in the following way 
	\begin{equation}\label{h-pars2}
		t_{1,2}=e^{-\alpha\pm \textup{i} \sigma_1},\quad t_{3,4}=e^{\alpha+\gamma-\eta \pm \textup{i} \sigma_2}, \quad
		t_{5,6}=e^{-\gamma\pm \textup{i} \sigma_3} \;.
	\end{equation}
	Then the contribution of chiral and vector multiplets to the superconformal index corresponds to the Boltzmann weights for the nearest-neighbor and the self-interaction, respectively. In the context of gauge/YBE correspondence, the spin-lattice model of Bazhanov and Sergeev can be identified with the four-dimensional $\mathcal N=1$ supersymmetric quiver gauge theory with $SU(2)$ gauge groups. On the vertices we have $SU(2)$ gauge group and the bifundamental matter content is represented as lines between gauge groups, see Fig 2. The partition function of the integrable model is equivalent to the superconformal index of the corresponding $SU(2)^N$ supersymmetric quiver gauge theory.

\section{The role of symmetries}
	
	Many properties of hypergeometric integrals have not yet been extended to the solutions of the star-triangle equation. In this section, we try to address some related issues.
	
	\subsection{Symmetry enhancement}
	
The star-triangle relation and the elliptic beta integral have the same number of parameters since we use the general $R$-charges (for rapidity parameters \cite{Yamazaki:2013nra,Yamazaki:2018xbx}) which gives three-parameter and the other three parameters come from the $SU(2)^3$ flavor symmetry. This is the case almost in all Ising-type models obtained via gauge/YBE correspondence. In terms of superconformal indices, the corresponding integral has a higher group of flavor symmetry. Therefore it is natural to ask may we obtain directly the elliptic beta integral type identity from integrable lattice models which has extended symmetry. 
	
	Let us consider the following integral 
	\begin{equation} \label{VSpir}
		Z(t_i;p,q) = \frac{(p;p)_{\infty} (q;q)_{\infty}}{2} \int_{{\mathbb T}}
		\frac{\prod_{j=1}^8 \Gamma(t_j z;p,q) \Gamma(t_j z^{- 1};p,q)}{\Gamma(z^{ 2};p,q) \Gamma(z^{- 2};p,q)}
		\frac{dz}{2 \pi i z}, 
	\end{equation}
	with the balancing condition $\prod_{j=1}^8 t_j=(pq)^2$. It has an integral transformation formula which enjoys a Weyl group symmetry of type $E_7$
	\begin{eqnarray} \label{starstar}
		Z (t_i;q)= Z(s_i;q)
		\prod_{1\leq j<k\leq 4} \frac{(q t_j^{-1}t_k^{-1},
			q t_{j+4}^{-1}t_{k+4}^{-1};q)_\infty}
		{(t_jt_k, t_{j+4}t_{k+4};q)_\infty},
	\end{eqnarray}
	where 
	\begin{equation}
		\left\{
		\begin{array}{cl}
			s_j =t_j \sqrt{\frac{q}{a_1a_2a_3a_4}},&   j=1,2,3,4  \\
			s_j = t_j \sqrt{\frac{a_5a_6a_7a_8}{q}}, &    j=5,6,7,8
		\end{array}
		\right. .
	\end{equation}
	Note that the map $t_j\to s_j$ is the key reflection extending the Weyl group $S_8$ (permutation group) of the root system $SU(8)$ to the Weyl group of the exceptional root system $E_7$. The integral (\ref{VSpir}) is a solution to the IRF-type Yang-Baxter equation, namely it is an $R$-matrix of the Bazhanov-Sergeev model \cite{Bazhanov:2011mz} and the identity (\ref{starstar}) gives the star-star relation. From field theory side this integral is the supersymmetric index of the four-dimensional ${\mathcal N}=1$ supersymmetric gauge theory with $SU(2)$ gauge group and $SU(8)$ flavor group; all matter multiplets in the fundamental representation of the gauge and flavor groups, vector multiplet is in the adjoint of the gauge group.
	
	Let us come back again to Spiridonov's elliptic beta integral. One can consider the beta integral (\ref{betaint}) as a limit of (\ref{VSpir}), then one can look for the  Weyl symmetry of the root system $E_6$ in this case. Actually, it is possible to realize $W(E_6)$ symmetry for the four-dimensional theory described in the previous section\footnote{The partition function of a theory with 
		flavor group $F$ has a Weyl group symmetry $W(F)$.} if it is coupled to a set of free five-dimensional hypermultiplets \cite{Dimofte:2012pd,Gahramanov:2013xsa}. In terms of superconformal index, one has the following expression
	\begin{align} \label{enhancedint}
		Z &=\prod_{1\leq i<j \leq 6} \frac{1}{\left((p q)^{\frac23}t_i^{-1} t_j^{-1};p,q\right)_{\infty}} \prod_{i=1}^6 \frac{1}{\left((p q)^{\frac13} t_i^{-1} w;p,q \right)_{\infty} \left((p q)^{\frac13} t_i^{-1} w^{-1};p,q \right)_{\infty}} \nonumber \\  
		& \quad \times \frac{(p,p)_\infty (q,q)_\infty}{2} \oint \frac{dz}{2 \pi i z} \frac{\prod_{i=1}^6 \Gamma(\sqrt[6]{p q}t_i z;p,q)\Gamma(\sqrt[6]{p q}t_i z^{-1};p,q)}{\Gamma(z^{2};p,q) \Gamma(z^{-2};p,q)} \; ,
	\end{align}
	where we used the double product notation $(z;p,q)_{\infty}=\prod_{i,j=0}^{\infty} (1-z q^i p^j)$.
	and redefined the fugacities $t_i \rightarrow (p q)^{-1/12} t_i$. The integral (\ref{enhancedint}) has an enhanced flavor symmetry $E_6$. One can show the presence of the extended symmetry by decomposition of the partition function into characters of
	the 
	flavor group which give sums of dimensions of irreducible representations of the $E_6$. This symmetry enhancement is a curious phenomenon which may take a place in terms of lattice spin models of statistical mechanics. There is a similar story for three-dimensional supersymmetric gauge theories \cite{Gahramanov:2015cva,Gahramanov:2016ilb,Gahramanov:2016wxi,Amariti:2018wht}. 
	
	\subsection{Modular properties and $SL(3,\mathbb{Z})$ symmetry}
	
	Our second remark is on the possible role of modular properties in integrable models. The footprints of modular invariance can be seen in many areas of mathematical physics, for instance, in string theory, it allows to complete of the perturbative sector with non-perturbative contributions, in two-dimensional conformal field theory it provides a tool for determining the spectrum of a theory. Modular invariance also appears in integrable lattice models in several ways, see e.g. \cite{Yang:1987mf,DiFrancesco:1991st,DiFrancesco:1989ha, Ercolessi:2013wg}, and the $SL(3,\mathbb{Z})$ symmetry is discussed in supersymmetric gauge theories, see, e.g. \cite{Spiridonov:2012ww,Brunner:2016nyk,Gadde:2020bov}.
	
	Here we will discuss the modular properties of elliptic gamma function. For our later purposes it is convenient to use the following reparametrization
	\begin{equation}
		p=e^{2 \pi i \tau}, \;\;\ q=e^{2 \pi i \sigma}, \;\; z=e^{2 \pi i u}.
	\end{equation}
	Then the elliptic gamma function has the following form
	\begin{equation}
		\Gamma(u;\tau, \sigma) \ = \ \prod_{i,j=0}^{\infty} \frac{1-e^{2 \pi i((1+j) \tau +(1+i)\sigma-u) }}{1-e^{2 \pi i (j\tau+i\sigma+u)}} \;,
	\end{equation}
	where $u, \sigma, \tau \in \mathbb{C}$ and $\text{Im} \tau, \text{Im} \sigma >0$. 
	
	Felder and Varchenko observed \cite{felder2000elliptic} that the elliptic gamma function enjoys an interesting transformation property 
	\begin{equation} \label{3gamma}
		\Gamma(
		\frac{u}{\sigma},\frac{\tau}{\sigma},-\frac{1}{\sigma})=
		e^{i\pi Q(u;\tau,\sigma)} {\Gamma(u,\tau,\sigma)}
		{\Gamma(
			{\frac{u-\sigma}{\tau},-\frac{1}{\tau},-\frac{\sigma}{\tau}})}
		\;,
	\end{equation}
	where $Q$ is the following polynomial
	\begin{align} \nonumber
		Q(u;\tau,\sigma)= \frac{u^3}{3\tau\sigma}
		-
		\frac{\tau+\sigma-1}{2\tau\sigma}
		u^2
		+
		\frac{\tau^2+\sigma^2+3\tau\sigma-3\tau-3\sigma+1}
		{6\tau\sigma}
		u \\ +
		\frac1{12}
		(\tau+\sigma-1)
		(\tau^{-1}+\sigma^{-1}-1).
	\end{align}
	This relation together with the following $q$-heat equation
	\begin{equation}
		\Gamma(u+\tau,\tau, \tau+\sigma) \Gamma(u,\tau+\sigma,\sigma) \ = \ \Gamma(u,\tau,\sigma) \;,
	\end{equation}
	imply that the elliptic gamma function has an $SL(3,\mathbb Z)$ symmetry transformation property. The polynomial $Q$ can be explicitly described in terms of the Bernoulli polynomial of the third order
	\begin{equation}
		B_{3,3}(u;\mathbb{\omega}) = \frac{1}{\omega_1\omega_2 \omega_3}
		\Bigl(u-\frac12\sum_{k=1}^3\omega_k\Bigr)\Bigl((u-\frac12\sum_{k=1}^3\omega_k )^2
		-\frac14 \sum_{k=1}^3\omega_k^2\Bigr)
	\end{equation}
	and the relation (\ref{3gamma}) gets the following form
	\begin{equation} \nonumber
		\Gamma(e^{2\pi i \frac{u}{\omega_2}};p,q)\Gamma(r e^{-2\pi i \frac{u}{\omega_1}};r, \tilde{q}) = \ e^{-\frac{\pi
				\textup{i}}{3} B_{3,3}(u;\mathbb{\omega})} \Gamma(e^{-2 \pi \textup{i}
			\frac{u}{\omega_3}};\widetilde{r},\widetilde{p}) \;,
	\end{equation}
	with the redefined parameters
	\begin{equation} \widetilde{p} \ = \ e^{-2 \pi \textup{i}
			\omega_2/\omega_3}, \quad \widetilde{q} \ = \ e^{-2 \pi \textup{i}
			\omega_2/\omega_1}, \quad \widetilde{r} \ = \ e^{-2 \pi \textup{i}
			\omega_1/\omega_3}
	\end{equation}
	\begin{equation} 
		p \ = \ e^{2 \pi \textup{i}
			\omega_3/\omega_2}, \quad q \ = \ e^{2 \pi \textup{i}
			\omega_1/\omega_2}, \quad r \ = \ e^{2 \pi \textup{i} \omega_3/\omega_1}.
	\end{equation}
	Let us introduce the modified gamma function  \cite{Spirtheta, Spiridonov:2012de}
	\begin{equation}
		{G}(u;{\bf \omega}) \ = \ e^{-\frac{\pi
				\textup{i}}{3} B_{3,3}(u;\mathbb{\omega})} \Gamma(e^{-2 \pi \textup{i}
			\frac{u}{w_3}};\widetilde{r},\widetilde{p}) \;.
	\end{equation}
	which is invariant under $SL(3,\mathbb Z)$ modular transformation. One can write an integral identity similar to the elliptic beta integral in terms of the modified elliptic gamma functions \cite{van2005unit}
	\begin{equation} \label{modbetaint}
		-\frac{(q;q)_\infty(p;p)_\infty(r;r)_\infty}{2(\tilde q;\tilde q)_\infty}
		\int_{-\omega_3/2}^{\omega_3/2} \frac{\prod_{n=1}^6 G(t_n+
			u;{\omega}) G(t_n-
			u;{\omega})} {G( 2u;{\omega}) G(- 2u;{\omega})} \frac{du}{\omega_2} = 
		\prod_{1\leq i<j\leq 6}G(t_i+t_j;{\omega}) \;,
	\end{equation} 
	with the balancing condition\footnote{To be more precise one needs also the conditions $\text{Im} (\omega_1/\omega_2)\geq 0$ and $\text{Im}
		(\omega_3/\omega_1), \text{Im} (\omega_3/\omega_2)>0$.} $\sum_{i=1}^6 t_i=\omega_1+\omega_2$ and $\text{Im}(g_i/\omega_3)<0$. The modified version of the elliptic beta hypergeometric integral satisfies the definition of elliptic hypergeometric integrals \cite{Spiridonovessay} and is well--defined for the case $|q| = 1$ and the integrand of this expression enjoys $SL(3, \mathbb{Z})$ modular property\footnote{There is an interesting observation made in \cite{Spiridonov:2012ww} that the 't Hooft anomaly matching conditions for dual theories are related to $SL(3,Z)$ modular transformation properties of the kernels of dual superconformal indices written as an integral over Coulomb branch moduli for a gauge group of the theory. }. One can construct a solution to the star-triangle relation in terms of modified elliptic gamma functions \cite{Spiridonov:2010em}
	\begin{equation}
		W_{\alpha}(\sigma_i,\sigma_j)  = G(e^{\alpha-\eta\pm i\sigma_i\pm i\sigma_j}; {\omega}) \;.
	\end{equation}
	Since this solution has an $SL(3,\mathbb{Z})$ modular invariance, one can use this additional symmetry in order to obtain some physical results. This remains an interesting guiding problem for future works. There is also an issue of whether one can obtain the identity (\ref{modbetaint}) from supersymmetric gauge theory computations. A typical mathematically-minded question: may we construct the whole class of solutions to the Yang-Baxter equation with an $SL(3,\mathbb{Z})$ modular symmetry? 
	
	In \cite{Yamazaki:2012cp,Kels:2015bda,Kels:2017vbc} various solutions to the Yang-Baxter equation were obtained in terms of the so-called lens gamma function, which is a combination of usual elliptic gamma functions. Modification of the lens gamma function solutions can be generalized in a similar fashion without any major difficulties.

	\subsection{Inversion relation}
	
	The inversion relation one of the main tools in integrable lattice models as the most direct path to solve models in statistical mechanics. The inversion relation was introduced by Stroganov for a certain two-dimensional vertex models on the square lattice \cite{Stroganov:1979et} and later used by many authors. Schematically, the inversion relations have the following form
	\begin{align}
		\SumInt_{\sigma}\,S(\sigma) W_{\eta-\alpha}(\sigma_i,\sigma) W_{\eta+\alpha}(\sigma,\sigma_j) & = \frac{1}{S(\sigma_i)}(\delta(\sigma_i\!+\!\sigma_j) + \delta(\sigma_i\!-\!\sigma_j))\,.
	\end{align}
	From the supersymmetric gauge theory side, the inversion relation corresponds to the chiral symmetry-breaking phenomenon for the partition function. The inversion relation also holds for some models which are not integrable (in terms of the Yang-Baxter equation), such as the Ising model in a magnetic field, the non-critical $q$-Potts model, etc. It would be interesting to understand the implications of this fact in the context of supersymmetric quiver gauge theories. 
	
	\subsection{Painleve equation}

	One compelling direction seems to be the establishment of further connections to the Painleve equations. It is possible to introduce $\tau$-functions in the theory of Painleve systems. The main goal of using them is that one can lift a representation of the affine Weyl groups on the level of the $\tau$-functions. As a result, one obtains various bilinear equations of Hirota type which are satisfied by these $\tau$-functions.
	
	There are some methods to obtain hypergeometric solutions for Painleve systems with large symmetries. In \cite{kajiwarareview,noumi2016} $\tau$-functions description is given for the elliptic Painleve equation of type $E_8^{(1)}$. The $q$-Painleve equation with affine Weyl group symmetry of type $E_8^{(1)}$ can be regarded as a limiting case of the elliptic one. In \cite{masudaE7,masudaE8} Masuda has proposed a method of constructing hypergeometric $\tau$-functions which are consistent with the action of the affine Weyl groups. One can use solutions of the star-triangle relation found in \cite{Gahramanov:2015cva,Gahramanov:2016ilb,Gahramanov:2017idz} in order to construct new hypergeometric solutions of the $q$-Painleve system with a certain Weyl group symmetry.
	
	It is known that discrete Painleve equations are integrable, see, e.g. \cite{takenawa2001algebraic}. Since the tau-functions for a certain discrete Painleve equation can be obtained from the star-triangle relation which defines the quantum integrability, there might be some relation between classical and quantum integrability.

	\section{Concluding remarks}
	
	The main focus of this short review is the recent developments in the study of integrable lattice models of statistical mechanics, mostly focused on aspects of research in which the author is involved. Instead of presenting some rigorous results, we concentrate on some mathematical issues which arise in the study of lattice models.

	Despite considerable progress, there are several important subjects connected with topics in this review that need further development. The realization of the gauge/YBE program is not finished and completely understood yet and still there remain many open questions.

	In this paper, we review integrable Ising-like square lattice models which have the property that the Boltzmann weights of the model depend only on the differences of the spin variables and rapidity variables. It would be interesting to obtain a model without such symmetries via the gauge/YBE correspondence.

	The main problem in statistical mechanics is to compute the partition function. Here we only discuss the integrability condition, namely if it is possible or not to calculate the partition function exactly, i.e. without any approximation. There are some techniques to compute the partition function for $N\rightarrow \infty$, i.e., in the thermodynamic limit. For instance, in \cite{Russo:2016ueu} the authors consider three-dimensional $\mathcal N=4$ QED\footnote{This theory is dual to $A_{N-1}$ (theory with $U(1)^{2N-1}$ gauge theory) quiver gauge theory \cite{Intriligator:1996ex,deBoer:1996mp,deBoer:1996ck}.} with $U(1)$ gauge group and with massive $N$ hypermultiplets on $S^3$ and compute the supersymmetric partition function in $N\rightarrow \infty$. In \cite{Dolan:2008qi} the authors computed the partition function for $\mathcal N=1$ theory with $SU(N)$ gauge group\footnote{By using their result in \cite{Bajc:2019vbp} there was found some relation to the cyclotomic polynomials. The cyclotomic polynomials also appear in the context of integrable models of statistical mechanics \cite{guttmann2005analytic,Guttmann2009}.}. It would be interesting to extend these computations to integrable models of statistical mechanics.

	\section*{Acknowledgements} The paper is based on talks given at the Department of Theoretical Physics, Steklov Mathematical Institute (Moscow, Russia), and at the Department of Mathematics, Izmir Institute of Technology (Izmir, Turkey). The work is originated from stimulating discussions with Deniz Bozkurt, Ege Eren, Mustafa Mullahasanoglu, and Shahriyar Jafarzade on the role of symmetries in lattice spin models, I thank them for their interest and suggestions. I learned much of the subject discussed here, in conversations with Vyacheslav Spiridonov, Hjalmar Rosengren, and Andrew Kels, I would like to thank them. The research on this project has received funding from the European Research Council (ERC) under the European Union's Horizon 2020 research and innovation program (QUASIFT grant agreement 677368) during the visit of the author to the Institut des Hautes Etudes Scientifiques (IHES), where some parts of this work were done. The work is partially supported by the Bogazici University Research Fund under grant number 20B03SUP3 and TUBITAK grant 220N106.

	\bibliographystyle{utphys} 
	\bibliography{references2}
\end{document}